\documentclass[aps,prd,preprintnumbers,nofootinbib,floatfix,floats,groupedaddress,twocolumn]{revtex4}

\usepackage{bm}
\usepackage[utf8]{inputenc}
\usepackage{latexsym}
\usepackage{dcolumn}
\usepackage{amsmath,amsfonts,amssymb}
\usepackage{graphicx,epsfig}
\usepackage{color}
\usepackage[active]{srcltx}
\usepackage[caption=false]{subfig}
\usepackage{slashed}
\usepackage{hyperref}
\usepackage{comment}
\usepackage{tikz}
\usetikzlibrary{shapes,snakes}
\usepackage{multirow}

\usepackage[mathscr]{euscript}
\usepackage{mathrsfs}
\usepackage{enumitem}
\def\nn{\nonumber}
\def\n{\nabla}
\def\pa{{\partial}}
\def\DM{\mathrm{d}}

\newcommand{\gae}{\lower 3pt \hbox{$\,\, \buildrel {\scriptstyle >}\over {\scriptstyle
\sim}\,\,$}}
\newcommand{\lae}{\lower 2pt \hbox{$\, \buildrel {\scriptstyle <}\over {\scriptstyle
\sim}\,$}}

\reversemarginpar

\def\eq#1{{Eq.~(\ref{#1})}}

\def\fig#1{{Fig.~\ref{#1}}}

\begin{document}

\title{Effect of tidal curvature on dynamics of accelerated probes 
}

 \author{Hari K}
 \email{harik@physics.iitm.ac.in}
 \author{Dawood Kothawala}
 \email{dawood@iitm.ac.in}
 \affiliation{Centre for Strings, Gravitation and Cosmology, Department of Physics, Indian Institute of Technology Madras, Chennai 600 036, India}

\date{\today}
\begin{abstract}
\noindent
We obtain a remarkable semi-analytic expression 
concerning the role of purely tidal curvature on accelerated probes, revealing some novel insights into the role of absolute vs. tidal acceleration in the response of such probes. The key quantity we evaluate is the relation between geodesic ($\tau_{\rm geod}$) and proper time ($\tau_{\rm acc}$) intervals between events on the probe trajectory. 
This is obtained as a covariant power series in curvature using a combination of analytical and numerical tools. A serendipitous observation then reveals that one can {\it exactly} sum all terms involving the {\it purely tidal} component ${\mathscr E}_n= R_{abcd} \varepsilon^{ab} \varepsilon^{cd}$ of curvature, with $\varepsilon^{ab}$ the bi-normal to the plane of motion:
$$
\tau_{\rm geod} = \frac{2}{\sqrt{{-\mathscr E}_n}} \sinh ^{-1}\Biggl[\sqrt{\frac{-{\mathscr E}_n}{a^2-{\mathscr E}_n}} \sinh \left(\frac{1}{2}  \sqrt{a^2-{\mathscr E}_n} \; \tau_{\rm acc}  \right) \Biggl]
$$
For classical clocks, the above result represents an interesting closed form contribution of tidal curvature to the differential ageing of twins in the classic {\it Twin paradox}. For quantum probes, it gives a thermal contribution to the {\it detector response} with a modified {\it Unruh temperature}
$$
[k_{\rm B} T]_{{\mathscr E}_n} = \frac{\hbar \sqrt{a^2- {\mathscr E}_n }}{2 \pi}
$$
As an operational tool, the computational framework we present and the corresponding results should find applications to a wide range of physical problems that involve measurements and observations by use of accelerated probes in curved spacetimes. 
\end{abstract}

\maketitle
\vskip 0.5 in
\noindent
\maketitle
\section{Introduction} \label{sec:intro} 
Physical effects in accelerated frames of reference have historically been a bedrock for some of the most important insights into classical as well as quantum physics. One often uses the results arrived at in accelerated frames in Minkowski spacetime to gain understanding of physical phenomena in a curved spacetime, gravitational time dilation and Hawking radiation being a couple of textbook examples which one can intuitively understand by studying accelerated clocks and quantum probes, respectively, in Minkowski spacetime (the latter, of course, if the well known Unruh effect). However, by its very nature, the principle of equivalence which one often uses to gain such an understanding does not give any insight whatsoever into the role of curvature in these phenomena. Curvature manifests itself through tidal acceleration, and this acceleration is to be ignored if one is to use the principle of equivalence. Moreover, for probes modelled as point systems, tidal forces can not possibly have any influence, and hence one might think that one need only worry about them when dealing with extended systems.  

In absence of any general, exact results concerning the effects of curvature, the above issues can only be justified through perturbative expansions in curvature, except in some very rare cases (such as (anti-)de Sitter spacetimes) where the full effect of the background constant curvature on accelerated probes can be evaluated.

In this paper, we show that, remarkably, one can capture in an analytic manner at least the effect of components of the Riemann tensor in the plane of motion of a uniformly accelerated trajectory. While there still remain terms which can only be described perturbatively, the subset of terms which can be summed to an analytic form yield some very important insights into the dynamics and response of accelerated probes in curved spacetimes. As a special case, we recover the well known results in (anti-)de Sitter spacetimes for which the parts that can not be summed vanish, and for a physically interesting class of motions in spherically symmetric spacetimes provided the derivatives of curvature are small, though the curvature itself might not be.

Our main result follows from the mathematical structure of the (square of) geodesic interval $\sigma(x, x')^2$ between points on an accelerated trajectory. We show that one can {\it exactly} sum all terms involving the {\it purely tidal} component ${\mathscr E}_n= R_{abcd} \varepsilon^{ab} \varepsilon^{cd}$ of curvature, with $\varepsilon^{ab}$ the bi-normal to the plane of motion. This summation gives the following analytic form for the contribution to $-\sigma(x, x')^2$
$$
\Biggl\{
\frac{2}{\sqrt{{-\mathscr E}_n}} \sinh ^{-1}\Biggl[\sqrt{\frac{-{\mathscr E}_n}{a^2-{\mathscr E}_n}} \sinh \left(\frac{1}{2}  \sqrt{a^2-{\mathscr E}_n} \; \tau_{\rm acc}  \right) \Biggl]
\Biggl\}^2
$$
where $\tau_{\rm acc}$ is the proper time along the accelerated curve. In Minkowski spacetime, the above expression reduces to it's well known Rindler form
$$
\sigma(x, x')^2 = - \Biggl\{ \frac{2}{a} \sinh \left(\frac{1}{2}  a \; \tau_{\rm acc}  \right) \Biggl\}^2
$$

We then show that the above expression has important implications when one considers detectors coupled of quantum fields in a curved spacetime. In Minkowski spacetime, as Unruh\cite{unruh-acc} has shown, such a detector will respond to the quantum fluctutations of the field. In particular, if the field is in Minkowski vacuum state, the quantum vacuum fluctuations as probed by an accelerated detector manifest as thermal fluctuations at a temperature
$$
k_{\rm B} T = \frac{\hbar a}{2 \pi}
$$
where $a$ is the acceleration. We will show that the contribution of ${\mathscr E}_n$ to the detector response is thermal part with temperature
$$
[k_{\rm B} T]_{{\mathscr E}_n} = \frac{\hbar \sqrt{a^2- {\mathscr E}_n }}{2 \pi}
$$
which again reduces to Unruh's result for Minkowski spacetime, but is now stated in a manner which is applicable to a spacetime for which ${\mathscr E}_n$ is not necessarily small.

The paper is structured as follows: In Sec \ref{sec:acc-motion}, we present a solution to the following mathematical problem: Given a curve $\mathcal{C}$ on a curved manifold, and a pair of points on $\mathcal{C}$, what is the relation between the arc-length distance between these points and the geodesic (``chordal'') distance between them, assuming a unique such geodesic exists? We present the set-up to find this relation to arbitrary orders of background curvature and acceleration of $\mathcal{C}$, and explicitly quote the result till $10^{\rm th}$ order in arc-length. In Sec \ref{sec:gen-cur-st}, we apply this result to hyperbolic motion in curved spacetime, and show that it admits a remarkable summation of certain terms yielding a closed form expression for them. In Sec \ref{sec:ms-proof}, a semi-analytic proof for the closed form expression in maximally symmetric spacetime is skecthed. In Sec \ref{sec:apps}, we use the above results in classical twin paradox, especially to (anti-)de Sitter and Schwarzschild spacetimes and to evaluate the clicking rate of the Unruh-deWitt detector, commenting also on the the role of the van Vleck determinant. Finally, we conclude by putting our results and analysis in a broader context and discussing possible generalizations.

\section{Accelerated vs Geodesic motion in curved spacetimes} \label{sec:acc-motion} 

\begin{figure}[htb!]%
    {{\includegraphics[width=0.3\textwidth]{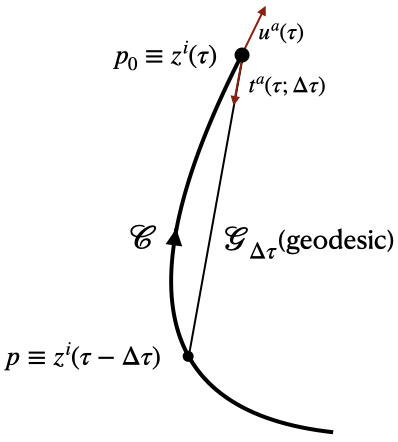} }}%
    \caption{The geometric setup for the problem. Two events on the hyperbolic trajectory $\mathcal{C}(\tau) \equiv z^i(\tau)$ are connected by a geodesic. The relation between the corresponding proper times
    holds the key information relevant for classical and quantum probes.}%
    \label{fig:trajectory}%
\end{figure}

Consider a curve $\mathcal{C}$ in an arbitrary spacetime, described by functions $z^i(\tau)$ in some coordinate system. Here, $\tau$ is the proper time parameter, $u^i(\tau)=\DM z^i/\DM \tau$ the four velocity, and $a^i = \nabla_{\bm u} u^i$ the acceleration. Consider two points $p_0, p$ separated by an interval $\Delta \tau$ along the curve, and assume $p$ to be in the geodesically convex normal neighbourhood of $p_0$. There will then be a unique geodesic $\mathcal{G}_{\Delta \tau}$ connecting $p_0$ and $p$ (with proper time parameter $\lambda$); see \fig{fig:trajectory}. We wish to find the relation between the proper length $\tau_{\rm geod}=\Delta \lambda$ of this geodesic and $\tau_{\rm acc}=\Delta \tau$. To do so, we set up Riemann normal coordinates (RNC) at $p_0$, with ${\widehat x}^i(p_0)=0$. Then, by definition of RNC\cite{poisson}, ${\widehat x}^a(p)=(\Delta \lambda){\widehat t}^a(0; \Delta \tau)$, where ${\widehat t}^a(0; \Delta \tau)$ is the tangent vector at $p_0$ with $\eta_{ab}{\widehat t}^{a}(0;\Delta\tau){\widehat t}^{b}(0;\Delta\tau)=-1$. 

To find the relation between $\Delta \lambda$ and $\Delta \tau$, we solve the equation for the trajectory ${\widehat z}^i(\tau)$ of $\mathcal{C}$ in RNC, 
\begin{eqnarray}
\frac{\DM^2 {\widehat z}^i}{\DM \tau^2} + 
{\widehat \Gamma}^i_{\phantom{i} bc} \frac{\DM {\widehat z}^b}{\DM \tau} \frac{\DM {\widehat z}^c}{\DM \tau} = {\widehat a}^i
\label{eq:geod-eq}
\end{eqnarray}
where ${\widehat a}^i$ the four acceleration, and ${\widehat \Gamma}^i_{\phantom{i} bc}(p_0)=0$ (property of RNC). Then, setting  $\tau=0$ at $p_0$, we can equate
\begin{eqnarray}
{\widehat z}^a(-\Delta \tau)={\widehat x}^a(p)=(\Delta \lambda)\; {\widehat t}^a(0; \Delta \tau)
\end{eqnarray}
to obtain
\begin{eqnarray}
(\Delta \lambda)^2 = \eta_{ab} {\widehat z}^a(-\Delta \tau) {\widehat z}^b(-\Delta \tau)
\label{eq:sigma-tau1}
\end{eqnarray}
To evaluate the LHS, we write a series solution of \eq{eq:geod-eq} as
\begin{eqnarray}
{\widehat z}^k(\tau) &=& \sum \limits_{n=0}^{\infty} \frac{\tau^n}{n!}
\left[ \frac{\DM^n {\widehat z}^k}{\DM \tau^n} \right]_{\tau=0}
\\
&=& {\widehat u}^k(0) \tau + {\widehat a}^k(0) \frac{\tau^2}{2} 
\\
&& \textcolor{white}{=} + 
\left( \dot{\widehat a}^k(0) - \dot{\widehat \Gamma}^a_{\phantom{a} bc}(0) {\widehat u}^b(0){\widehat u}^c(0) \right) \frac{\tau^3}{6}
\ldots
\label{eq:geod-eq-series-sol}
\end{eqnarray}
where the derivatives on the RHS are computed by successive differentiations of \eq{eq:geod-eq}. The argument ``$0$'' above means everything is evaluated at $p_0$, and the overdot indicates differentiation along the trajectory. To clean up the above expression, we express the proper time derivatives of $\dot{\widehat a}^k$ in terms of covariant derivatives along ${\widehat u}^k$; for example, $\dot{\widehat a}^k = {\widehat \nabla}_{\bm {\widehat u}}{\widehat a}^k - {\widehat \Gamma}^k_{\phantom{k} ij} {\widehat u}^i {\widehat u}^j$. with ${\widehat \nabla}_{\bm {\widehat u}} \equiv {\widehat u}^i {\widehat \nabla}_i$. 
This introduces more Christoffel symbol terms, making the computation a laborious task that is best done using a symbolic package. To simplify 
analysis, we will make the following assumptions:

(i) The motion is hyperbolic, and 

(ii) Derivatives of curvature, $\nabla_i R_{abcd}$, are ``small". Note that we are not assuming the curvature itself to be small, only that it changes slowly in the domain of interest.

The above assumptions are reasonable given that we are interested in obtaining leading curvature corrections to uniformly accelerated systems; we will discuss the relaxations of these in the final section. Even with these assumptions, the calculation is computationally expensive, particularly since, as we will show, the leading curvature corrections themselves appear only at $O(\tau^6)$! All our numerical computations have been done using \texttt{CADABRA}\cite{cadabra-kp-lb}. 

We now give the mathematical simplifications that arise under the above assumptions:

\underline{\textbf{(i) \textit{Hyperbolic motion}}}: The mathematical description of uniformly accelerated hyperbolic motion in curved spacetime was first given by Rindler\cite{rindler}, essentially by imposing the Serret-Frenet formulae\cite{book-synge} to spacetime curves. The hyperbolic motion is described by the equations
\begin{eqnarray}
{\nabla}_{\bm u}u^k &=& a n^k
\hspace{.25cm}; \hspace{.25cm} {\nabla}_{\bm u} a = 0
\\
{\nabla}_{\bm u}n^k &=& a u^k
\label{eq:rindler-conds}
\end{eqnarray}
where $a^2=a_i a^i$ and $n^i=a^i/a$. These conditions are stated more elegantly in terms of the Fermi derivative
$$
\frac{D_F}{\DM \tau}  \bm u = \frac{D_F}{\DM \tau}  \bm n = 0
$$
where, by definition, 
$\frac{D_F}{\DM \tau} \bm v = \nabla_{\bm u} \bm v + (\bm v \cdot \bm u) \bm a - (\bm v \cdot \bm a) \bm u$.
This has important implication that $\bm n$ can be used as one of the basis vectors while describing $\mathcal{C}$ in Fermi normal coordinates based on it. This, in turn, means that one can use $a$ as one of the spatial coordinates, a fact familiar from its many appearances in the study of thermal properties of static horizons.

The Rindler conditions immediately help us take care of all covariant derivatives of $a^i$ along $u^i$ in \eq{eq:geod-eq-series-sol}, since they imply
\begin{eqnarray}
({\nabla}_{\bm u})^p a^k &=& a^{2p} u^k \hspace{0.825cm} (p=1,3,5, \ldots)
\\
({\nabla}_{\bm u})^p a^k &=& a^{2p-1} n^k \hspace{0.5cm} (p=2,4,6, \ldots)
\label{eq:acc-derivs}
\end{eqnarray}

\underline{\textbf{(ii) \textit{Ignoring derivatives of Riemann}}}: If one ignores all the derivatives of Riemann, the metric in RNC can be expressed as \cite{Katanaev}:
\begin{eqnarray}
{\widehat g}_{ab} &=& \eta_{ab} + \frac{1}{2}\sum \limits_{k=1}^{\infty} (-1)^k \frac{2^{2k+2}}{(2k+2)!} \mathbb{M}^{m_1}_{a} \mathbb{M}^{m_2}_{m_1} .... \mathbb{M}^{m_{(k-1)}}_{b} 
\end{eqnarray}
where the matrix $\mathbb{M}$ is defined by
\begin{eqnarray}
{\mathbb M}^i_j = R^i_{\phantom{i}ajb} {\hat x}^a {\hat x}^b   
\end{eqnarray}
It is easy to identify the series above as related to the series expansion of $\sin x/x$, yielding
\begin{eqnarray}
{\widehat g}_{ab} &=& [{\sin{\mathbb{M}}}/{\mathbb{M}}]^i_{(a} \eta_{b)i} 
\label{eq:metric-rnc}
\end{eqnarray}
This also gives a closed form expression for the van Vleck determinant $\Delta(0, {\hat x})$ in RNC, which we will need later:
\begin{eqnarray}
\Delta(0, {\hat x}) = \sqrt{{\rm det}[{\mathbb{M}}/{\sin{\mathbb{M}}}]}
\label{eq:vvd}
\end{eqnarray}
This follows from the fact that, in RNC, $\Delta(0, {\hat x})=1/\sqrt{-{\hat g}}$.
\\
\\
Using the assumptions (i) and (ii) above, we obtain
\begin{equation}
\begin{aligned}
\dot{\widehat{z}^i}(0)
&=u^i \rvert_{p_0} = u^i(0)
\\
\ddot{\widehat{z}^i}(0)
&=\left[ an^i-\Gamma^i_{m k}u^m u^k\right]_{p_0}=\left[a n^i \right]_{p_0} 
\\
\dddot{\widehat{z}^i}(0) 
&= \left[a\,u^k \pa_k n^i - \Gamma^i_{m j, k}u^k u^m u^j -2 \Gamma^i_{m j}(u^k\pa_ku^j)\right]_{p_0} \\
&= \left[a\,\n_{\bm u} n^i - 3 \Gamma^i_{k m}u^ka^m - \Gamma^i_{m j,k}u^k u^m u^j + \Gamma^i_{m j}\Gamma^{j}_{k l}u^m u^k u^l \right]_{p_0}
\\
&= \left[a\,\n_{\bm u} n^i - \Gamma^i_{m j,k}u^k u^m u^j \right]_{p_0}
\end{aligned}
\end{equation}
Higher derivatives can be similarly obtained. We now have all the tools to explicitly evaluate \eq{eq:sigma-tau1}. Before giving the results for arbitrary curved spacetime, we give the {\it exact} results for maximally symmetric spacetimes.

We must mention that there exists an alternate route to the above derivation using the properties of Synge's World function (one-half of the squared geodesic distance) as given in Ref. \cite{book-jls}. However, this method becomes increasingly difficult at higher order. Our method based on RNC, in conjunction with \texttt{CADABRA}, is easier to implement.

\subsection{Maximally symmetric spacetimes}

For maximally symmetric spacetimes, the Riemann tensor is $R_{abcd}=\Lambda (g_{ac} g_{bd} - g_{ad} g_{bc} )$, with $\Lambda$ constant. It is more convenient to work with the so called embedding coordinates $X^i$, with $Z^i(\tau)$ representing $\mathcal{C}$ in these coordinates (for convenience, we denote all quantities in embedding coordinates by capitalised versions of their corresponding symbols in RNC). The metric is given by \cite{book-weinberg} 
\begin{eqnarray}
g_{ab} = \eta_{ab} + 
		\frac{\Lambda }{1-\Lambda \eta_{ij} X^i X^j} \eta_{ac} \eta_{bd} X^c X^d
\end{eqnarray}
and the Christoffel connections are $\Gamma^a_{\phantom{a} bc} = \Lambda X^a g_{bc}$. Therefore, $\Gamma^a_{\phantom{a} bc} U^b U^c=-\Lambda X^a$ and $\Gamma^a_{\phantom{a} bc} U^b A^c=0$, which enormously simplifies the conditions \eq{eq:rindler-conds}. They become
\begin{eqnarray}
{\dot U}^k &=& A^k + \Lambda X^k
\\
{\dot A}^k &=& A^2 U^k
\\
{\ddot U}^k &=& \left( A^2 + \Lambda \right) U^k
\end{eqnarray}
from which one can easily show that
\begin{eqnarray}
\left(\frac{\DM}{\DM \tau}\right)^{2p+1} A^k &=& a^2 q^{2p} \; U^k \hspace{0.825cm} (p=0,1,2, \ldots)
\nn \\
\left(\frac{\DM}{\DM \tau}\right)^{2p \phantom{+1}} A^k &=& a^2 q^{2(p-1)} \; (A^k+\Lambda X^k)
\hspace{0.5cm} (p=1,2,3, \ldots)
\nn
\label{eq:acc-derivs-max-symm}
\end{eqnarray}
where $q = \sqrt{a^2 + \Lambda}$. Plugging this into \eq{eq:geod-eq-series-sol}, the trajectory is easily obtained as
\begin{eqnarray}
Z^k(\tau) = \frac{\sinh q \tau}{q} U^k(0) + \frac{\cosh q \tau - 1}{q^2} A^k(0)
\end{eqnarray}
The corresponding trajectory in RNC is then obtained by the mapping between RNC and embedding coordinates (see Ref \cite{hk-dk} for a proof): 
\begin{equation}
    X^i = \Delta^{-1/{(D-1)}} {\hat x}^i
\end{equation}
where $\Delta=\Delta (0,{\hat x})$ is the van Vleck determinant, and $D$ the dimension of spacetime. This, when plugged into \eq{eq:sigma-tau1}, finally gives (after some simplifications) the relation between $\tau_{\rm geod}=\Delta \lambda$ and $\tau_{\rm acc}=\Delta \tau$
\begin{eqnarray}
\tau_{\rm geod}^2 &=& \frac{1}{\Lambda} 
\Biggl[
\sin^{-1}
\left\{
\sqrt{- \Lambda \tau_{\rm geod, \eta}^2 (1+\frac{1}{4}\Lambda \tau_{\rm geod, \eta}^2) } 
\right\}
\Biggl]^2
\nn \\
\\
&=& \frac{4}{\Lambda}
\Biggl[
\sinh ^{-1}
\left\{\sqrt{\frac{\Lambda}{q^2}} \sinh \left(\frac{q \, \tau_{\rm acc} }{2} \right) \right\}
\Biggl]^2
\label{eq:sigma2-max-symm}
\\
\tau_{\rm geod, \eta}^2 &=& \frac{4}{q^2} \sinh^2\left({\frac{q \tau_{\rm acc}}{2}}\right)
\end{eqnarray}
\\
(The second equality above follows from standard identities associated with hyperbolic functions.) 
Note that 
$$
\lim \limits_{\Lambda \to 0} \tau_{\rm geod}^2 = \frac{4}{a^2} \sinh^2\left({\frac{a \tau_{\rm acc}}{2}}\right)
$$
\\
which is the familiar expression for Rindler trajectories in Minkowski spacetime. In fact, $\tau_{\rm geod, \eta}^2$ has precisely the form corresponding to Rindler motion in Minkowski spacetime, but with acceleration $q=\sqrt{a^2 + \Lambda}$. 

With future anticipation, we highlight the periodicity
\begin{equation}
\tau_{\rm geod}^2 \left[\tau_{\rm acc}\right] = \tau_{\rm geod}^2 \left[\tau_{\rm acc} + (2 \pi i) {q^{-1}}\right] 
\end{equation}
This periodicity is key to thermal properties that a probe, put on an accelerated trajectory, will assign to the quantum field which is in inertial vacuum state.
\\
\section{General curved spacetimes} \label{sec:gen-cur-st}

For general spacetimes, even with assumptions (i) and (ii), the computation becomes unwieldy pretty fast, and we therefore employ \texttt{CADABRA} for the same. To $O((\Delta \tau)^{10})$, the relation between $\Delta \lambda=\tau_{\rm geod}$ and $\Delta \tau=\tau_{\rm acc}$ is finally obtained (after a laborious computation) as:
\begin{widetext}
\begin{align}
\tau_{\rm geod}^2 &={\tau^{2}_{\rm acc}} + \frac{1}{12}a^{2} {\tau^{4} _{\rm acc}} 
+ \frac{1}{360}\left(a^{4} + 3 a^2 {\mathscr E}_n \right) {\tau^{6}_{\rm acc}} \nn
+ \frac{1}{20160}\left(a^{6} + 17 a^2 R_{\bullet 0 n 0} R^{\bullet}\,_{0 n 0} + 18 a^{4} {\mathscr E}_n \right) {\tau^{8}_{\rm acc}}
		\\
& + \frac{1}{1814400} \left(a^{8} + 81 a^{6} {\mathscr E}_n + 675 a^{4} R_{\bullet 0 n 0} R^{\bullet}\,_{0 n 0} + 336 a^4 R_{\bullet n n 0} R^{\bullet}\,_{n n 0} 
+ 155 a^2 R^{\bullet}\,_{0 n 0} R_{\bullet 0 \bigstar 0} R^{\bigstar}\,_{0 n 0} \right) {\tau^{10}_{\rm acc}} + O(\tau^{12}_{\rm acc})
\\
 &=
\tau^{2}_{\rm acc} 
+ \frac{1}{12}a^{2} {\tau}^{4}_{\rm acc} + \frac{1}{360}\left( a^{4} + 3 a^2 {\mathscr E}_n \right) {\tau}^{6}_{\rm acc} 
+ \frac{1}{20160}\left(a^{6} + 17 a^2 {\mathscr E}_n^2 + 18 a^{4} {\mathscr E}_n \right) {\tau}^{8}_{\rm acc}  \nn
\\
& 
+ \frac{1}{1814400} \left( a^{8} + 81 a^{6} {\mathscr E}_n +  339 a^{4} {\mathscr E}_n^2 + 155 a^2 {\mathscr E}_n^3 \right) {\tau}^{10}_{\rm acc}
+ O(\tau^{12}_{\rm acc})
+ \mathscr{R}_A
\label{eq:sigma-tau-relation}
\end{align}
\end{widetext}
where, for better readability, we have used following convenient notations: the symbols ``$0$", ``$n$" on Riemann indicates contraction of the corresponding indices with $u^i, n^i$, with ${\mathscr E}_n = R_{0n0n}=R_{abcd} u^a n^b u^c n^d$ and so on. Indices which are summed over are indicated by $\bullet, \bigstar, $etc. The first equality is our first main result, whose extensions and generalizations we will discuss in the last section. The second equality represents a remarkable structure following from the first one. To arrive at it, we extract all the terms in the first equality which depend {\it purely} on ${\mathscr E}_n$ by expanding the summations; for instance, $R_{\bullet n n 0} R^{\bullet}\,_{n n 0} = - {\mathscr E}_n^2 + R_{A n n 0} R^{A}\,_{n n 0}$, with $A$ representing an index in the space orthogonal to $u^i$ and $n^i$. Hence, $\mathscr{R}_A$ collectively represents all terms that have at least one Riemann tensor with at least one index which is neither $0$ nor $n$. Our main observation then follows from the power series representation of the following analytic function:
\begin{widetext}
\begin{eqnarray}
\Biggl[
\frac{2}{\sqrt{r_0}}
\sinh^{-1}
\left\{
\sqrt{\frac{r_0}{1+r_0}} \sinh \left(\frac{1}{2} \sqrt{1+r_0} w\right)
\right\}
\Biggl]^2
= 
w^2 &+& \frac{w^4}{12} + \frac{1}{360} (1-3 r_0) w^6 + \frac{\left(17 r_0^2-18 r_0+1\right) w^8}{20160}
\nonumber \\
&+& \frac{\left(-155 r_0^3+339 r_0^2-81 r_0+1\right) w^{10}}{1814400}+O\left(w^{12}\right)
\end{eqnarray}
\end{widetext}
which, on comparison with \eq{eq:sigma-tau-relation}, yields
\begin{eqnarray}
\tau_{\rm geod} &=& \frac{2}{\sqrt{{-\mathscr E}_n}} 
\sinh ^{-1}
\Biggl[\sqrt{\frac{-{\mathscr E}_n}{a^2-{\mathscr E}_n}} \sinh \left(\frac{\sqrt{a^2-{\mathscr E}_n} \, \tau_{\rm acc} }{2} \right) \Biggl] 
\nonumber \\
&& \hspace{.25cm} + \mathscr{R}_A
\label{eq:sigma2-general}
\end{eqnarray}

We must highlight that the above expression is exact, and it is very much possible that the terms represented by $\mathscr{R}_A$ also admit some sort of summation to a closed form; we have, however, not been able to do so as yet, and will comment on it in the final section.

\textit{\bfseries Spherically symmetric spacetimes:}

Remarkably, our results also cover the important class of spherical symmetric spacetimes, with metric: 
\begin{eqnarray}
\DM s^2 = -f(r)\DM t^2 + h(r) \DM r^2 + r^2 \DM \Omega^2
\end{eqnarray}
where $f(r)$ and $h(r)$ are arbitrary functions, and $\DM \Omega^2$ is the line element of the $2-$sphere. Due to spherical symmetry, and the maximally symmetric $2-$sphere part, the Riemann tensor of this class of spacetimes has a unique decomposition, as can be easily shown by a direct generalization of the form given for Schwarzschild metric in Ref. \cite{candelas}. For arbitrary choice of $f(r)$ and $h(r)$, the Riemann tensor is decomposed as,
\begin{eqnarray}
R_{\textsf{abcd}}&=&\frac{ff^{\prime}h^{\prime}+h\left[f^{\prime 2}-2ff^{\prime\prime}\right]}{4f^2h^2}\left(g_{\textsf{ac}}g_{\textsf{bd}}-g_{\textsf{ad}}g_{\textsf{bc}}\right) \nn\\
R_{ABCD}&=&\frac{h-1}{r^2h}\left( g_{AC}g_{BD}-g_{AD}g_{BC} \right) \nn\\
R_{\textsf{a}A\textsf{b}B}&=&\frac{1}{2rh}\;g^{\prime}_{\textsf{ab}}g_{AB}\nn
\end{eqnarray}
where, indices $\textsf{a},\textsf{b}\in \{t,r\}$, indices $A,\,B\in \{\theta,\phi\}$ and prime is the derivative with respect to the radial coordinate $r$. 

In these spacetimes, consider hyperbolic motion in the $t-r$ plane; then the unit vectors $u^i$ and $a^i$ are in $t-r$ plane. In order to have a non-zero value for $\mathscr{R}_A$, there must be at least one Riemann tensor component in $t,r,A$ direction, which do not exists. A more formal proof of this is given in the next section, for a maximally symmetric spacetimes. The same argument, however, works in spherical symmetry as well. This will eventually imply ${\mathscr R}_A=0$ and hence give
 \begin{eqnarray}
\tau_{\rm geod} &=& \frac{2}{\sqrt{{-\mathscr E}_n}} 
\sinh ^{-1}
\Biggl[\sqrt{\frac{-{\mathscr E}_n}{a^2-{\mathscr E}_n}} \sinh \left(\frac{\sqrt{a^2-{\mathscr E}_n} \, \tau_{\rm acc} }{2} \right) \Biggl] \nn \\
&+& \text{terms involving $\nabla R_{abcd}$}
\label{eq:sigma2-spherical}
\end{eqnarray}
For the special case of static observers at $r=R, \theta, \phi=$ constant (and $g(r)=1/f(r)$), ${\mathscr E}_n = f''(R)/2|$. For (A-)dS spacetimes, with $f(r)=1-\Lambda r^2$, this reproduces the exact result derived earlier, while for Schwarzschild/Reissner-Nordstrom, the above gives a distinct contribution which, to the best of our knowledge, has not appeared in the literature. If there exists a radius at which $a$ diverges, this will lead approximately to the Rindler result in Minkowski spacetime, but at an arbitrary radius, the additional curvature term might become significant. 

\section{A Semi-analytic proof} \label{sec:ms-proof}

The result above was obtained from numerical computations in \texttt{CADABRA}, combined with the observation that the coefficients match exactly with the power series expansion of certain combination of hyperbolic functions. This combination is by no means trivial, and it is only through a fortuitous circumstance that we were able to arrive at it. However, one notes that the final expression, \eq{eq:sigma2-general}, is similar to the expressions \eq{eq:sigma2-max-symm} obtained in the case of maximal symmetry, apart from the $\mathscr{R}_A$ term. This allows us to give the following semi-analytic proof for our expression.

From \eq{eq:metric-rnc} and the conditions \eq{eq:rindler-conds}, it is easy to see that terms in $\tau_{\rm geod}$ which are higher order in Riemann appear in the form $R_{a b c \bullet}R^{\bullet}\,_{g h k} \ldots R^{\bigstar}\,_{d e f}$ with the free indices dotted with $n^{a}$'s and $u^{a}$'s. Consider such a term with $k+1$ factors on Riemann tensors. We write this expression in a general form as 
\begin{eqnarray}
{\bar Q}_{abcdef} &=& R_{a b c m_1} Q^{m_1}_{m_k}[k-1] R^{m_k}\,_{d e f}
\nn \\
{\rm where \;\;\;\;} Q^{m_1}_{m_k}[k-1] &=& \prod^{k-1}_{i=1}R^{m_i}\,_{p_{(2i-1)} p_{2i} m_{(i+1)}} 
\nn
\end{eqnarray}
(All the $p_i$'s are dotted with $u^a$ or $n^a$, and hence we do not display them on $Q^a_b[k-1]$.)
This tensor has the property 
$Q^{m_1}_{m_k}[k-1]=R^{m_1}\,_{p_1 p_2 m_2} Q^{m_2}_{m_k}[k-2]$ which is obvious from above. Now, in maximal symmetry, due to the structure of Riemann, one has the following fact: since $(a,b,c,d,e,f,p_i)\in \left\{0,n\right\}$, we must also have $m_1, m_k \in \left\{ 0,n\right\}$, otherwise ${\bar Q}_{abcdef}=0$. Therefore, for $R^{m_1}\,_{p_1 p_2 m_2}$, the indices, $m_1, p_1, p_2 \in \{0,n\}$, which in turn implies that $m_2 \in \{0,n\}$ in maximally symmetry. Since the tensor $Q^{m_1}_{m_k}[k-1]$ obeys the above recursion relation, all other product inside this tensor will be the same form. 

Therefore, we have shown that in maximal symmetry, the degeneracy arising because of the fact that $R^{A}\,_{n n A} = R^{A}\,_{0 0 A} = R^{n}\,_{0 0 n} = \Lambda$ is broken in presence of acceleration, and terms other than $R_{0n0n}$ {\it do not contribute to the series}. We have therefore proved that in a generic curved spacetime, all the terms in $\tau_{\rm geod}$ which have all the Riemann tensor indices belonging to $0$ or $n$, must sum to the expression in maximal symmetry with $\Lambda \to (-R_{0n0n})$.

\section{Applications} \label{sec:apps} 

\subsection{The classical twin paradox}

The mapping of our problem to the classical version of the twin paradox is immediate from \fig{fig:trajectory}, read as follows: The twins separate at $p_1$, with one of them moving on the hyperbolic trajectory $\mathcal{C}$ by using some external force, and the other following a geodesic. Provided the twins meet, the expression we have derived directly gives the relation between their ages. This, of course, is a somewhat unrealistic setup since the relative Lorentz boost $\gamma_{\rm rel}=-\bm u \cdot \bm t \neq 1$, implying a non-zero relative velocity when they meet. One can presumably remedy this by smoothing the accelerated trajectory near the meeting points, which would involve introducing a mild time dependence in $a$, with only a tiny modification to the above expression.

The twin paradox problem for (anti)de Sitter spacetime is addressed in Ref. \cite{twin-1,Ads}, and the case of Schwarzschild spacetime is taken up in Ref. \cite{twin-2}. However, these references work with specific accelerated trajectories, while our result is more general and applicable to any hyperbolic trajectory, with $a$ arbitrary.

\textit{de Sitter spacetime}: 
Consider the twins, one accelerating with constant acceleration and other one stationary in de Sitter spacetime. The hyperbolic trajectory of the accelerated twin has been discussed by Rindler \cite{rindler}, and has been used in Ref. \cite{twin-1} to discuss the twin paradox in de Sitter. The trajectory, in standard FLRW coordinates, is given by
\begin{eqnarray}
t(\tau_{\rm acc})=\frac{1}{H}\ln\left(\frac{\Psi}{2q^2e^{q\tau_{\rm acc}}}\right);\quad r(\tau_{\rm acc})=\frac{a\left(e^{q\tau_{\rm acc}}-1\right)^2}{\Psi}\nn\\
\end{eqnarray}
where, $H$ is the Hubble constant, $\Psi$ is defined as $\Psi=H(q+H)e^{2q\tau_{\rm acc}}+2a^2e^{q\tau_{\rm acc}}-H(q-H)$ and $q$
is same as defined before, with $H^2=\Lambda$. The geodesic distance in de Sitter can be found using $\tau_{\rm geo}^2=H^{-2} \cos^{-1}\left[\left({\eta^2+\eta\prime^2-\ell^2}\right)/{2\eta\eta\prime}\right]^2$ where $\eta$ is the conformal time and $\ell^2$ is the squared spatial distance. It is not difficult to verify, by writing the trajectory in terms of conformal time and substituting in the equation for geodesic distance, we recover \eq{eq:sigma2-max-symm} as expected. The simplicity of the expressions presented we have derived, vis a vis the more cluttered ones when restricting to special coordinate systems, is worth highlighting. The geometric insights it provides is an added bonus.

\begin{figure*}[!ht]
\includegraphics[width=0.8\textwidth]{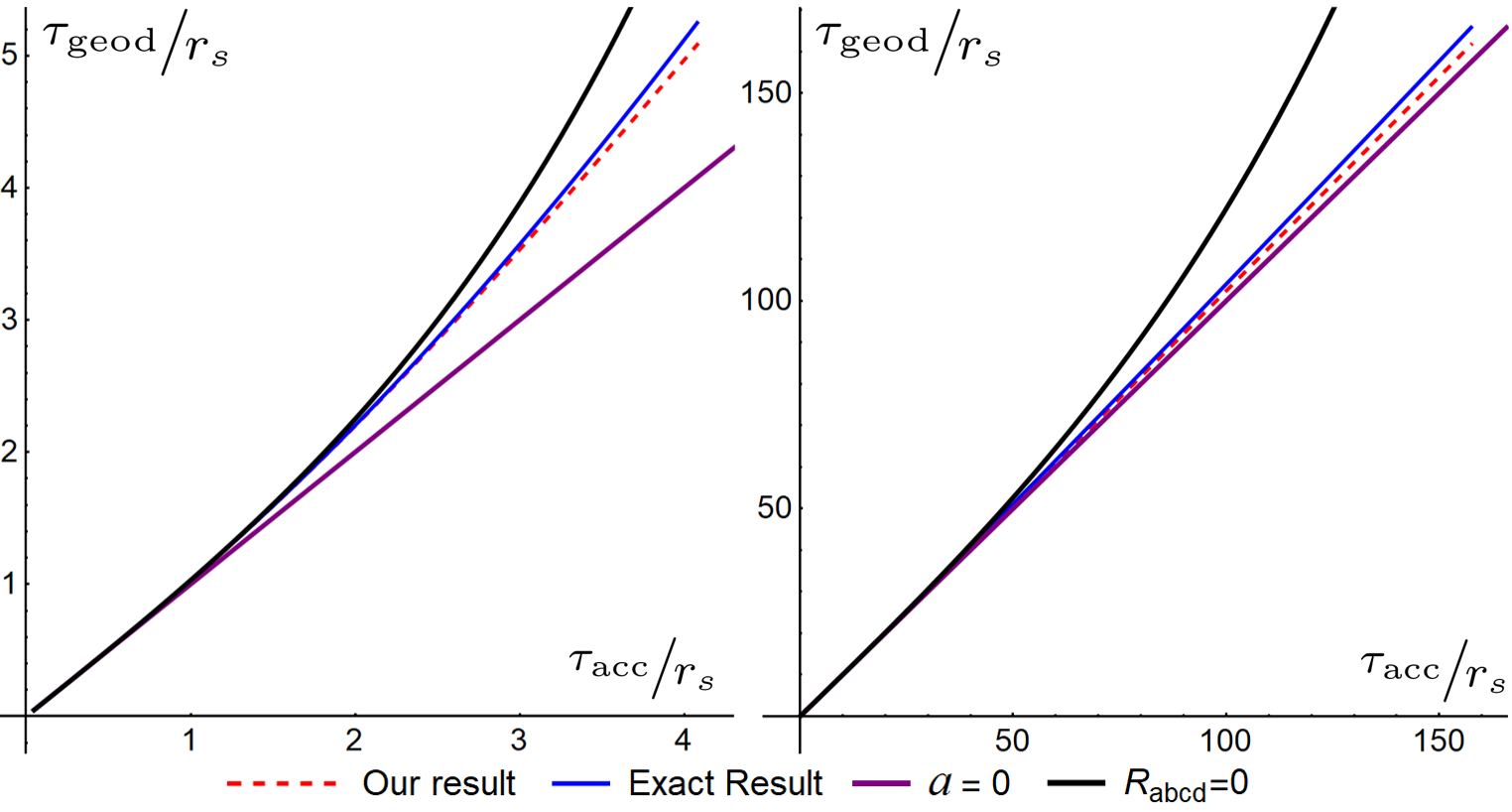}
\caption{Comparing the relation between the $\tau_{\rm acc}$ and $\tau_{\rm geod}$ for for static observer and radial geodesic observers in Schwarzschild ($r_s=2GM/c^2$). The parameters involved are the radius of the static observer, $r_0$, and radius of the turning point for the geodesic observer, $r_{\rm tp}$. Left plot is for $r_0=1.2 r_s, r_{\rm tp}=0.5 r_0$, and the right one for $r_0=5 r_s, r_{\rm tp}=2 r_0$}
\label{fig:comparison}
\end{figure*}

\textit{anti - de Sitter spacetime}:
For anti-de Sitter spacetime, $\Lambda<0$ and some interesting features emerge. The negative curvature constant will impose a bound to the value of $\tau_{\rm acc}$. This can be demonstrated using \eq{eq:sigma2-max-symm}. Assume $a>\sqrt{|\Lambda|}$,
then, for $\Lambda<0$, the RHS of \eq{eq:sigma2-max-symm} will have $\sinh^{-1} \to \sin^{-1}$ and $\tau_{\rm acc}$ will be bounded due to the bound on the range of function. Thus, for $a>\sqrt{|\Lambda|}$, the proper time of acceleration will have a bound as,
\begin{eqnarray}
\tau_{\rm acc} &\leq& \frac{2}{\sqrt{a^2-|\Lambda|}}
\sinh ^{-1}
\Biggl[\sqrt{\frac{a^2-|\Lambda|}{|\Lambda|}} \Biggl] 
\leq \frac{2}{\sqrt{|\Lambda|}}
\nn
\end{eqnarray}
It is easy to verify that the upper bound here is consistent with the fact that the maximum geodesic length in anti-de Sitter spacetime is ${\pi}/{\sqrt{|\Lambda|}}$ \cite{Ads}.

\textit{Schwarzschild}: In the case of Schwarzschild geometry, the simplest scenario would be to consider motion in $t-r$ plane, and take for the accelerated twin the one at rest at $r=r_0, \theta, \phi=$ constant. The second twin can then be set on the radial geodesic that goes out from $r=r_0$, turns around, and reaches back to $r=r_0$ to meet the static twin. Fixing the values for $r_0$ and the turning point one can relate the $\tau_{\rm geod}$ and $\tau_{\rm acc}$ \cite{twin-2}. The \fig{fig:comparison} shows the comparison of this, as given, for example, in  \cite{twin-2}, with the result we have obtained. The deviations of our result from the exact one is due to derivatives of the Riemann tensor, which our formula does not capture. This is evident when the accelerated observer is very near to the event horizon, $r_s={2GM}/{c^2}$ of the black hole and the turning point of the geodesic is very large. The derivatives of the Riemann tensor will matter for such instances. For a stationary observer at $r \gg r_0$, our result is closer to the exact relation.

\subsection{Quantum probes - the Unruh-de Witt detector}

In the case of Minkowski spacetime, the Unruh-Dewitt detector is coupled to a real scalar field in the Minkowski vacuum. The detector is accelerated and the transition probability of the detector to detect particles in the accelerated frame of the Minkowski vacuum is obtained. This will give a planckian spectrum. A broad discussion of this can be found in Ref. \cite{book-birrell}. For generalising this a curved spacetime, we consider the detector coupled to a real scalar field $\phi$ in an arbitrary Hadamard state. The Hadamard state in arbitrary curved spacetime is equivalent to the Minkowski vacuum in the Minkowski spacetime. Instead of considering the interaction from past infinity to future infinity, the switching function\cite{ud-det-1} is used to limit the interaction to take place for some finite time. The interaction Hamiltonian for the detector is given by $H_{int}(\tau)=c\,m(\tau)\chi(\tau)\phi(x(\tau))$
where $\tau$ is the proper time along the accelerated trajectory, $c$ is the coupling constant, $m$ is the detector's monopole moment and $\chi(\tau)$ is the switching function. Due to the presence of the switching function, the interaction only takes place when $\chi$ is positive definite. The joint system of detector being in the ground state and particle in arbitrary Hadamard state is given by, $\left|\sf H\right\rangle\otimes\left|0\right\rangle$. Then using the perturbation theory, the probability of the detector in $\left|1\right\rangle$ is given by,
$
\mathcal{P}=c^2 \left| \left\langle 0 \right| m(0) \left|1 \right\rangle \right| ^2 \mathcal{F}(\omega)
$
where, $\omega$ is the difference in energy between the states and $\mathcal{F}(\omega)$ is the response function given by
\begin{equation}
\mathcal{F}(\omega)=\int^{\infty}_{-\infty}\,\DM \tau \int^{\infty}_{-\infty}\, \DM \tau^{\prime\prime}\, \chi(\tau) \chi(\tau^{\prime\prime})e^{-i\omega\left(\tau-\tau^{\prime\prime}\right)}G^{+}(\tau^{\prime\prime},\tau)
\end{equation}
where $G^{+}$ is the Wightmann two-point function of the field. The straightforward way to account for the switching is to use a window function and a simplest way of doing this is choosing $\chi(\tau)$ as,
\begin{equation}
\chi(\tau)=\Theta(\tau-\tau_0)\Theta(\tau^{\prime}-\tau)
\end{equation}
where, $\tau_0$ is the proper time when the detector is switched on and $\tau^{\prime}$ is the proper time when detector is read (at the base point). Using this window function, the response function becomes,
\begin{equation}
\mathcal{F}(\omega)=\int^{\tau^{\prime}}_{\tau_0}\,\DM \tau \int^{\tau^{\prime}}_{\tau_0}\, \DM \tau^{\prime\prime}\, e^{-i\omega\left(\tau-\tau^{\prime\prime}\right)}G^{+}(\tau^{\prime\prime},\tau)
\end{equation}
Using the coordinate transformation $u=\tau,\,s=\tau-\tau^{\prime\prime}$ for $\tau^{\prime\prime}<\tau$, $u=\tau-\tau^{\prime\prime},\, s=\tau$ for $\tau<\tau^{\prime\prime}$ and the property of Wightmann function  $G^{+}(\tau,\tau^{\prime\prime})=G^{+^{*}}(\tau^{\prime\prime},\tau)$, one can obtain the response function as in Ref. \cite{ud-det-2},
\begin{equation}
\mathcal{F}(\omega)=2\int^{\tau^{\prime}}_{\tau_0}\,\DM u \int^{u-\tau_0}_{0} \DM s\, \text{Re}\left[e^{-i\omega s}G^{+}(u,u-s)\right]
\end{equation}
The more physically significant quantity is the transition rate which defined as the time derivative of $\mathcal{F}(\omega)$ given by,
\begin{equation}
\dot{\mathcal{F}}(\omega)=2\int^{\tau^{\prime}-\tau_0}_{0} \DM s\, \text{Re}\left[e^{-i\omega s}G^{+}(\tau^{\prime},\tau^{\prime}-s)\right]
\end{equation}
The Wightmann two-point function for the Hadamard state will be given by, $G^{+}(x,x^{\prime}):=\left\langle \sf H \middle|\hat{\phi}(x) \hat{\phi}(x^{\prime}) \middle| \sf H \right\rangle$ where $x$ is the coordinate position. The projection of this Wightmann function along the accelerated trajectory $G^{+}(\tau,\tau^{\prime})=G^{+}(x,x^{\prime})$, is taken to find the transition rate. The Wightmann two-point function in a Hadamard state is given by\cite{wightmann-fn}, 
\begin{equation}
G^{+}(x,x^{\prime})=\frac{1}{4\pi^2}\left(\frac{\Delta^{\frac{1}{2}}(x,x^{\prime})}{\sigma_{\epsilon}^2(x,x^{\prime})} + v(x,x^{\prime})\ln\left[\sigma_{\epsilon}^2(x,x^{\prime})\right] \right)
\end{equation}
where $\Delta(x,x^{\prime})$ is the van Vleck determinant, $\epsilon$ is a small positive parameter, $\sigma_{\epsilon}^2(x,x^{\prime})$ is the Synge world function\cite{book-jls} with $i\epsilon$ prescription given as, $\sigma_{\epsilon}^2(x,x^{\prime}):=\sigma^2(x,x^{\prime})+2i\epsilon\left[ T(x)-T(x^{\prime}) \right]$, where $T(x)$ is increasing global time function and $v(x,x^{\prime})$ is polynomial function of $\sigma^2(x,x^{\prime})$.

Thus the pull-back of the Wightmann function onto the accelerated trajectory which is the detector worldline, requires the relation between the proper time $\tau$ on the worldline and the squared geodesic distance $\sigma^2(x,x^{\prime})$ between the considered time interval. The effect due to the form of window function and the regularisation are discussed more in Refs. \cite{ud-det-1,ud-det-2}. We will be focusing on the effects due to curvature in $\sigma^2(x,x\prime)$. Using \eq{eq:vvd} and our result,
\begin{eqnarray}
\frac{\Delta^{\frac{1}{2}}(x,x^{\prime})}{\sigma^2(x,x^{\prime})}
&=& 
\frac{ - {\mathscr E}_n \left(\sqrt{{\rm det}[{\mathbb{M}}/{\sin{\mathbb{M}}}]}\right)^{1/2}}
{4 \Biggl(
\sinh ^{-1}
\Biggl[\sqrt{\frac{-{\mathscr E}_n}{a^2-{\mathscr E}_n}} \sinh \left(\frac{\sqrt{a^2-{\mathscr E}_n} \, \tau_{\rm acc} }{2} \right) \Biggl]
\Biggl)^2}
\nn \\
&+& {\mathscr R}_A + \textrm{terms depending on $\nabla R_{abcd}$}
\end{eqnarray}
The pole structure of the response function will not be affected by $\Delta(0, {\hat x})$, since $\Delta(0, {\hat x})=1+O(\tau_{\rm acc}^2)$.
A computation similar to the one for Rindler motion in Minkowski spacetime then yields
\begin{eqnarray}
\dot{\mathcal{F}}(\omega; \tau) &=& \frac{\omega}{2 \pi} \Biggl[{\exp \left(\frac{\hbar \omega}{[k_{\rm B} T]_{{\mathscr E}_n}}\right)-1}\Biggl]^{-1} 
\nn \\
&+& 
(\textrm{terms depending on ${\mathscr R}_A, \nabla R_{abcd}$})
\end{eqnarray}
with
\begin{eqnarray}
[k_{\rm B} T]_{{\mathscr E}_n} = \frac{\hbar}{2 \pi} \sqrt{a^2- {\mathscr E}_n }
\end{eqnarray}
The above result is exact for (anti) de-Sitter spacetimes, with ${\mathscr E}_n=-\Lambda$ since $\sigma^2(x,x^\prime)$ is exact. In the last section, we will elucidate the geometrical connection between our result and the well known derivations based on global embeddings (GEMS). Note that the above thermal contribution can also be understood directly in terms of the identity $\tau_{\rm geod}^2 \left[\tau_{\rm acc}\right] = \tau_{\rm geod}^2 \left[\tau_{\rm acc} + (2 \pi i) {q^{-1}}\right]$ satisfied by our result to the leading order, ignoring $\mathscr{R}_A$. This immediately implies the periodicity in Euclidean time of the two-point function  $G^{+}(x,x^\prime)$ (to this order), implying a temperature $k_BT={\hbar q}/{2\pi}$  \cite{ud-temp, ud-ds1, ud-ds2}. 

Apart from (A-)dS, ${\mathscr R}_A$ vanishes for several other spacetimes of physical significance for specific hyperbolic motion. Two relevant examples are:

\underline{1. \it Spherically symmetric spacetimes}: For static observes in spherically symmetric spacetimes with $g^{rr}=-g_{00}=f(r)$ in standard coordinates $(t, r, \theta, \phi)$, we have ${\mathscr R}_A=0$ and ${\mathscr E}_n=f''(r)/2|_{p_0}$, and the above result becomes
\begin{eqnarray}
[k_{\rm B} T]_{{\mathscr E}_n} = \frac{\hbar}{2 \pi} \sqrt{a^2 - \frac{f''(r)}{2} \Biggl|_{p_0} } 
+ O(\nabla R_{abcd})
\end{eqnarray}

\underline{2. \it FLRW spacetimes}: For hyperbolic motion in FLRW universes ($k=0$) with scale factor $a_{1}(t)$, once again we have ${\mathscr R}_A=0$ and ${\mathscr E}_n=-{\ddot a}_1(t)/a_1(t)|_{p_0}$.
\begin{eqnarray}
[k_{\rm B} T]_{{\mathscr E}_n} = \frac{\hbar}{2 \pi} \sqrt{a^2 + \frac{{\ddot a}_1(t)}{a_1(t)} \Biggl|_{p_0} } 
+ O(\nabla R_{abcd})
\end{eqnarray}

There might many more examples of hyperbolic trajectories in well-known spacetimes for which ${\mathscr R}_A=0$ due to symmetry, in which case our result captures the full Riemann tensor modification modulo any effects due to derivatives of Riemann.

Let us also comment on the $a \to 0$ limit of the final result. For $-{\mathscr E}_n>0$, this leaves a finite temperature proportional to $a_0 = \sqrt{-{\mathscr E}_n}$, whose interpretation remains unclear, since there is no natural choice for $\bm n$ when $a=0$. Note, however, that the difference $\sqrt{a^2+a_0^2}-a_0$ has the right limits; it goes as $a^2/2a_0$ for $a \ll a_0$, and as $a$ for $a \gg |a_0|$. It is not clear to us if this difference has any physical relevance. 

We end this section with a comment on a couple of papers (Refs. \cite{afshordi}) which give a ``prescription" to attribute temperature to quantum field theories in curved spacetime based on the structure of the two-point function. Essentially, the curvature correction in this prescription arises by Taylor expanding the van Vleck determinant in the numerator of the two-point function to $O(\tau_{\rm acc}^2)$, and combining it with the pure acceleration term of the same order in the denominator. The prescription then ``reads-off" an effective temperature $\sqrt{a^2 - R_{ab} u^a u^b + w}$, where $w$ is the contribution from the state dependent term in the two point function. This is clearly different from the temperature that can be identified from the thermal part of the detector response, although both yield the same result for maximally symmetric spacetimes. In computing the detector response, it is clear that the van Vleck determinant does not contribute at all to the coincidence limit poles of the two-point function, neither does it contribute to the residues (see also \cite{ud-det-1}). The only divergence(s) in $\Delta(x,y)$ arise(s) at the caustics, and these are infrared divergences.


\section{Implications and Discussion} \label{sec:implications} 
Measurements and observations in the accelerated frames are tools to understand some of the most subtle aspects of classical and quantum physics such as the notion of {\it inertia}, Mach principle, the {\it equivalence principle}, Unruh effect. The general rule of thumb, when using the results derived in accelerated frames in Minkowski spacetime to deduce aspects of gravitational field, is the following: As long as $a \gg \sqrt{|R_{abcd}|}$, where $|R_{abcd}|$ is typical magnitude of the curvature tensor, one expects an accelerated frame of reference to mimic effects of a gravitational field to the lowest order. Even though such a condition seems to make sense, it is too stringent since one is ignoring the fact that acceleration $a^i$ defines a spacelike direction, and hence components of Riemann tensor in the $\bm u \bm \wedge \bm n$ plane might lead to interesting effects, and even non-trivial, effects. Very often, when one talks about effects such as, say, the Unruh effect, in curved spacetime, one often imagines the additional curvature contribution coming because of modification to {\it dynamical equations}, such as the d' Alembartian and hence $G^+(x, x')$. To the best of our knowledge, the dependence on curvature arising due to {\it kinematical} aspects of the trajectory is hardly ever considered. In more precise terms, the curvature dependence inherent in the relation
$$
\sigma^2(z^i(\tau), z^i(\tau')) \to \sigma^2(\tau'-\tau, R_{abcd})
$$
is hardly ever discussed. This dependence is independent of the dynamical equations in the problem one is studying and, as we have shown here, is crucial as it contributes in an important, non-trivial manner to standard results. Another way of proceeding with the computation would have been to use identities associated with Synge's World function bi-scalar \cite{book-jls}; however, this method becomes cumbersome beyond fourth order, and, at this order, one does not encounter any curvature terms at all.

A hint towards the importance of such a kinematical term comes from (anti) de Sitter spacetime, wherein a temperature of $\sqrt{a^2 + \Lambda}$ can be associated with hyperbolic motion in a precise sense -- an exact result with no restriction on relation $a$ and $|\Lambda|$. In this paper, we have shown that an analog of such a result holds in an arbitrary curved spacetime. As illustrated in the paper, the geodesic interval between two points on a hyperbolic curve can be expressed as a series in which all the terms involving components of Riemann tensor in the plane of motion, ${\mathscr E}_n=R_{abcd} u^a n^b u^c n^d$, sum up exactly to an analytic function. Remarkably, this function depends on acceleration only through the parameter $q=\sqrt{a^2 - {\mathscr E}_n }$. Being exact, such a dependence eliminates any need for a restriction on relative magnitudes of $a$ and ${\mathscr E}_n$, thereby allowing one to read-off the interesting {\it non-pertubative} effects arising from the combination. While one still does not have any handle on off-the-plane components of Riemann, and must therefore assume them to be small compared to $\{a^2, {\mathscr E}_n\}$, the analytic dependence on ${\mathscr E}_n$ already leads to insights which the Taylor series itself would never have lead to. 

\textit{Connection with GEMS}: For (A)dS, our result is exact and reproduces the the famous result by Deser and Levin \cite{ud-ds2} for (anti) de-Sitter spacetimes considered as embedded sub-manifolds of a five dimensional flat spacetime (the so called GEMS approach). We here briefly highlight the mathematical connection between these results. We start with the Gauss-Codazzi equation,
\begin{eqnarray}
R_{\textsf{ABCD}} h^{\textsf A}_{a}h^{\textsf B}_{b}h^{\textsf C}_{c}h^{\textsf D}_{d}=R_{abcd}+\varepsilon \left(K_{ad}K_{bc}-K_{ac}K_{bd}\right)
\end{eqnarray}
where, $h^{\textsf{A}}_{a}$ is the projector, $K_{ab}$ is the extrinsic curvature, $\varepsilon=\pm 1$ which depend on the magnitude of unit normal on the hypersurface and indices $\textsf{A,B,C,D}$ indicate the higher dimensional space. Since the embedding is into a higher dimensional flat space(time), we have $R_{\textsf{ABCD}}=0$, and hence
\begin{eqnarray}
R_{abcd}=\varepsilon \left(K_{ac}K_{bd}-K_{ad}K_{bc}\right)
\end{eqnarray}
Moreover, for the standard embedding of (anti) de Sitter spacetimes in $D=5$, we have $K_{ab}=-K_{uu} g_{ab}$ with $K_{uu}:=K_{ab}u^au^b$. This immediately implies 
$R_{abcd}u^a n^b u^c n^d = - \varepsilon K_{uu}^2$. Now, we use the following result from differential geometry (see Ref. \cite{dk-duality} for a pedagogical proof)
\begin{eqnarray}
a_5^2=a^2+\varepsilon K_{uu}^2
\end{eqnarray}
where, $a_5$ is the acceleration of $u^a$ in $5-d$ flat spacetime, $a$ is it's acceleration in the $4-d$ spacetime. The above identities immediately yield
\begin{eqnarray}
a_5^2=a^2 - R_{abcd}u^a n^b u^c n^d = q^2
\end{eqnarray}
thereby providing a connection with our result. A similar analysis can be done for spherically symmetric spacetimes, but then the embedding spacetime has dimension greater than $5$. For a $6$ dimensional embedding of Schwarzschild, for instance, one would have to compute both the extrinsic curvatures (corresponding to the two normals) to establish a similar connection. 

We conclude with a few comments that we hope will be relevant for future work:

\begin{enumerate}[wide, labelwidth=!, labelindent=0pt]
    \item In deriving our results, we have ignored the derivatives of Riemann, but the method we have described can be readily extended to incorporate these derivatives. However, we do not expect any exact results to arise out of such a study, nor will the result established here be affected by these additional terms. Our expressions are exact for (A)dS, and provides a remarkably good fit to the known cases of hyperbolic motion in Schwarzschild; \fig{fig:comparison} shows the comparison for trajectories with same magnitude of acceleration in different spacetimes.

\begin{figure}[ht!]
\centering
\includegraphics[width=0.4\textwidth]{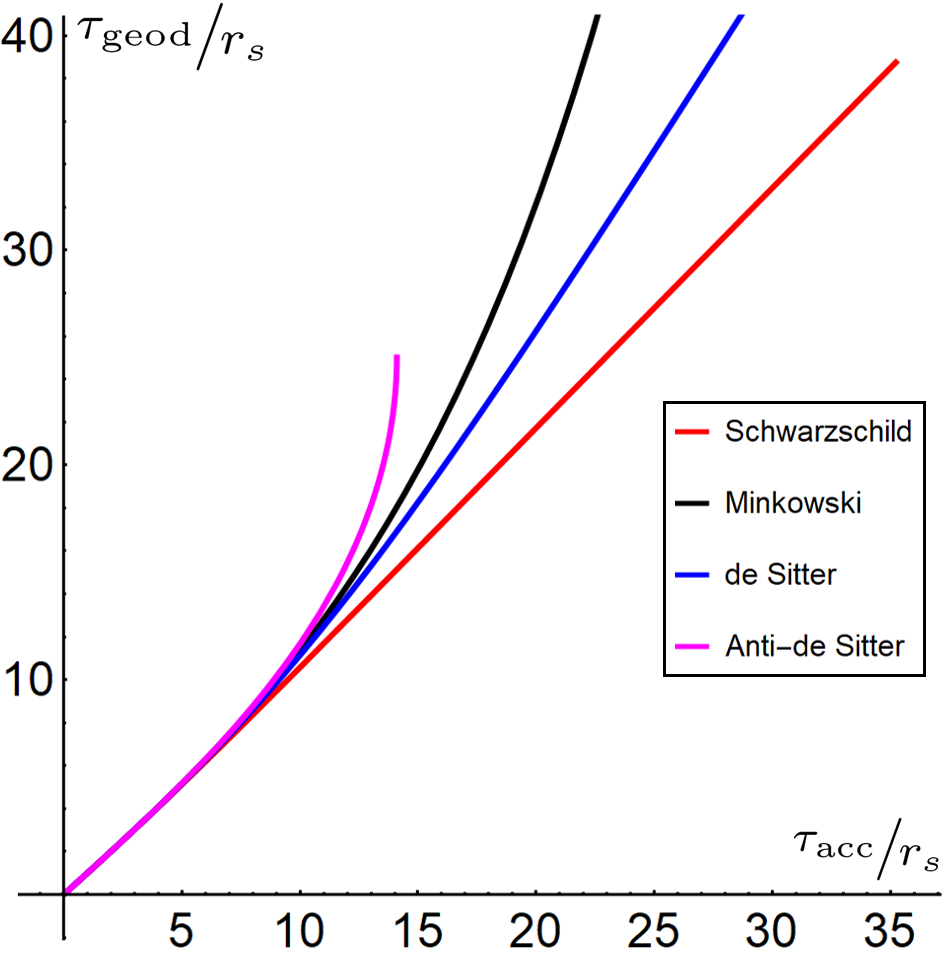}\hfill
\caption{Comparing the relation between the $\tau_{\rm acc}$ and $\tau_{\rm geod}$ for different spacetimes. The values chosen are, $r_0=2r_s$, $r_{\rm tp}=2r_0$ for Schwarzschild and $|\Lambda|={a^2}/{2}$ for de Sitter and anti-de Sitter where $a$ is the magnitude of acceleration for Schwarzschild observer.}
\label{fig:comparison}
\end{figure}

\item It would be nice to apply our result to spherically symmetric black hole solutions in particular, and check how the resultant temperature there is related to the Hawking temperature. In this context, a couple of remarks can be made: First, if we consider static observers near a Killing horizon, then $a$ diverges and hence dominates over $\mathscr{E}_n$. The temperature we have obtained will then give the red-shifted Hawking temperature to leading order. Second, we know that the Unruh effect relies crucially on Lorentz invariance (LI). Ref. \cite{viqar-louko} explicitly discusses this issue by pointing out the non-trivial consequences that may arise from violations of LI, by analyzing atoms modelled as UD detectors. In this context, it is worth noticing that the analytic function appearing on the RHS of \eq{eq:sigma2-general} represents the full contribution of the 
boost invariant part of Riemann with respect to boosts in the $\bm u \wedge \bm n$ plane. To see this, note that the boost weight zero components of Riemann are $R_{0n0n}, R_{0nAB}, R_{A[0n]B}$ and $R_{ABCD}$. The last three do not appear in the expression for $\tau_{\rm geod}$ for hyperbolic motion. 
Hence, the contribution we have extracted is also the purely boost invariant 
part of Riemann tensor. It would nice to see if any further insights can be gained from our result concerning detectors in the background of a black hole emitting Hawking radiation \cite{smerlak}.

\item Our analysis brings into sharp focus the role of absolute vs. tidal acceleration in the response of accelerated probes, and it therefore will hopefully provide a first step towards quantifying the qualitative picture often associated with vacuum phenomenon such as Schwinger, Unruh, or Hawking effects -- that of {\it peeling} of trajectories caused by a classical background field (electric or gravitational).

\item The analysis/results presented here can be generalised in several directions. The most important generalisation would be to obtain a summed version for the contribution of $\mathscr{R}_A$.
Beyond this, one might generalize further by including derivatives of curvature as well as deviations from hyperbolic motion. However, we do not expect any analytic results in such cases for arbitrary trajectories and/or curvature. The case of time dependent acceleration is non-trivial even in Minkowski spacetime, with no closed form expressions available. In our view, a much more insightful procedure would be to use the method described in this paper to generalise to curved spacetime the known cases of other stationary trajectories in Minkowski spacetime, in particular the case of uniform rotation. The conditions for hyperbolic motion will now need to be modified to incorporate rotation. Although this will complicate the computations, such an extension might yield useful insights into behavior of rotating detectors \cite{rotating-atom}. Any closed form expression for this case will surely also bring in information about additional components of Riemann.
\end{enumerate}

\acknowledgements
HK would like to thank IIT, Madras and Ministry of Human Resources and Development (MHRD), India for financial support.
        
\widetext


\begin{thebibliography}{99}

%
\bibitem{unruh-acc}
S. A Fulling, \href{https://journals.aps.org/prd/abstract/10.1103/PhysRevD.7.2850}{Phys. Rev. D {\bf 7}, 2850} (1973); P. C. W. Davies, \href{https://iopscience.iop.org/article/10.1088/0305-4470/8/4/022}{J. Phys. A: Math. Gen. {\bf 8} 609} (1975); W. G. Unruh, \href{https://journals.aps.org/prd/abstract/10.1103/PhysRevD.14.870}{Phys. Rev. D {\bf 14}, 870} (1976)

\bibitem{poisson}
E. Poisson, A. Pound, and I. Vega, \href{https://link.springer.com/article/10.12942%2Flrr-2011-7}{Living Rev. Relativity {\bf 14}, 7} (2011), \href{https://arxiv.org/pdf/1102.0529.pdf}{arXiv:1102.0529 [gr-qc]}.

\bibitem{cadabra-kp-lb}
K. Peeters. \href{https://doi.org/10.1016/j.cpc.2007.01.003}{Computer Physics Communications {\bf 176}, 550} (2007), \href{https://arxiv.org/abs/hep-th/0701238}{arXiv:hep-th/0701238}; L. Brewin, \href{https://www.sciencedirect.com/science/article/pii/S0010465509003415}{Comput. Phys. Commun {\bf 181} 489} (2010), \href{https://arxiv.org/pdf/0903.2085.pdf}{	arXiv:0903.2085 [gr-qc]}; L. Brewin, \href{https://iopscience.iop.org/article/10.1088/0264-9381/26/17/175017}{Classical Quantum Gravity {\bf 26}, 175017} (2009), \href{https://arxiv.org/pdf/0903.2087v2.pdf}{arXiv:0903.2087 [gr-qc]}

\bibitem{rindler}
W. Rindler, \href{https://journals.aps.org/pr/pdf/10.1103/PhysRev.119.2082}{Physical Review, {\bf 119}, 2082} (1960)

\bibitem{book-synge}
J. L. Synge and A. Schild, {\it Tensor Calculus} (University of Toronto Press, Toronto, 1949)

\bibitem{Katanaev}
M. O. Katanaev, \href{https://link.springer.com/article/10.1134/S199508021803006X}{Lobachevskii J. Math. \bf{39}, 464} (2018)

\bibitem{book-jls}
J. L. Synge, {\it Relativity: The General Theory} (North Holland publishing company, Amsterdam, 1960)

\bibitem{book-weinberg}
S. Weinberg, {\it Gravitation and cosmology: Principles and applications of the general theory of relativity} (Wiley, New York, 1972).

\bibitem{hk-dk}
K. Hari and D. Kothawala, \href{https://journals.aps.org/prd/abstract/10.1103/PhysRevD.101.124066}{Phys. Rev. D {\bf 101}, 124066} (2020), \href{https://arxiv.org/pdf/2003.10169.pdf}{arXiv:2003.10169 [gr-qc]}

\bibitem{candelas}
P. Candelas, \href{https://journals.aps.org/prd/pdf/10.1103/PhysRevD.21.2185}{Phys. Rev. D {\bf 21}, 2185} (1980).

\bibitem{twin-1}
S. Boblest, T Müller, and G Wunner, \href{https://iopscience.iop.org/article/10.1088/0143-0807/32/5/001/meta}{European journal of physics {\bf 32}, 1117} (2011), \href{https://arxiv.org/pdf/1009.3427v1.pdf}{arXiv:1009.3427 [gr-qc]}

\bibitem{Ads}
L. M. Soko{\l}owski \href{https://www.worldscientific.com/doi/abs/10.1142/S0219887816300166}{Int. J. Geom. Methods Mod. Phys {\bf 13}, 1630016} (2016), \href{https://arxiv.org/pdf/1611.01118.pdf}{arXiv:1611.01118 [gr-qc]}

\bibitem{twin-2}
L. M. Soko{\l}owski \href{https://link.springer.com/article/10.1007/s10714-012-1337-4}{Gen. Relativ. Gravit {\bf 44}, 1267} (2012), \href{https://arxiv.org/pdf/1203.0748.pdf}{arXiv:1203.0748 [gr-qc]}; {\O}. Gr{\o}n and S. Braeck \href{https://link.springer.com/article/10.1140/epjp/i2011-11079-7}{Eur. Phys. J. Plus {\bf 126} 79}  (2011), \href{https://arxiv.org/pdf/0909.5364.pdf}{arXiv:0909.5364 [gr-qc]}

\bibitem{book-birrell}
N. D. Birrell and P.~C.~W.~Davies, \textsl{Quantum Fields in Curved Space}, Cambridge University Press, Cambridge (1982).

\bibitem{ud-det-1}
J. Louko and A. Satz, \href{https://iopscience.iop.org/article/10.1088/0264-9381/25/5/055012}{Class. Quant. Grav. {\bf 25}, 055012} (2008), \href{https://arxiv.org/pdf/0710.5671.pdf}{arXiv:0710.5671 [gr-qc]}; A. Satz, \href{https://iopscience.iop.org/article/10.1088/0264-9381/24/7/003}{Class. Quant. Grav. {\bf 24}, 1719} (2007), \href{https://arxiv.org/pdf/gr-qc/0611067.pdf}{arXiv:gr-qc/0611067}

\bibitem{ud-det-2}
S. Schlicht, \href{https://iopscience.iop.org/article/10.1088/0264-9381/21/19/011}{Class. Quant. Grav. {\bf 21} 4647} (2004), \href{https://arxiv.org/pdf/gr-qc/0306022v2.pdf}{arXiv:gr-qc/0306022}

\bibitem{wightmann-fn}
M. J. Radzikowski, \href{https://link.springer.com/content/pdf/10.1007/BF02100096.pdf}{ Commun. Math. Phys. {\bf 179}, 252} (1996); B. S. Kay and R. M. Wald, \href{https://www.sciencedirect.com/science/article/pii/037015739190015E}{Phys. Rept. {\bf 207}, 49} (1991)


\bibitem{ud-temp}
J S Dowker, \href{https://iopscience.iop.org/article/10.1088/0305-4470/10/1/023/pdf}{J. Phys. A: Math. Gen. {\bf 10} 115} (1977)

\bibitem{ud-ds1}
H. Narnhofer, I. Peter and W. Thirring, \href{https://doi.org/10.1142/S0217979296000611}{Int. J. Mod. Phys. B {\bf 10} 1507}(1996).

\bibitem{ud-ds2}
S. Deser and O. Levin, \href{https://iopscience.iop.org/article/10.1088/0264-9381/14/9/003}{Class. Quant. Grav. {\bf 14} L163} (1997), \href{https://arxiv.org/pdf/gr-qc/9706018.pdf}{arXiv:gr-qc/9706018}; S. Deser and O. Levin, \href{https://journals.aps.org/prd/pdf/10.1103/PhysRevD.59.064004}{Phys. Rev. D {\bf 59}, 064004} (1999), \href{https://arxiv.org/pdf/hep-th/9809159.pdf}{	arXiv:hep-th/9809159}

\bibitem{afshordi}
M. Lynch, N. Afshordi, 
\href{https://iopscience.iop.org/article/10.1088/1361-6382/aae792}{Class. Quant. Grav. {\bf 35} 225008 (2018)}
\href{https://arxiv.org/abs/1611.06619}{arXiv:1611.06619};
A. Dhumuntarao, J. Ghersi, N. Afshordi, 
{Report No: SCG-2018-04}
\href{https://arxiv.org/abs/1804.05382}{arXiv:1804.05382};

\bibitem{dk-duality}
D. Kothawala, \href{https://aapt.scitation.org/doi/10.1119/1.3553231}{Am. J. Phys. {\bf 79} 624} (2011), \href{https://arxiv.org/pdf/1010.2238.pdf}{arXiv:1010.2238 [physics.class-ph]}

\bibitem{viqar-louko}
Viqar Husain, Jorma Louko, 
\href{https://journals.aps.org/prl/abstract/10.1103/PhysRevLett.116.061301}{Phys. Rev. Lett. {\bf 116}, 061301} (2016)
\href{https://arxiv.org/abs/1508.05338}{arXiv:1508.05338}


\bibitem{smerlak}
Some discussion on the possible role of curvature in the response of detectors in certain model spherically symmetric spacetimes can be found in: 
M. Smerlak and S Singh, \href{https://journals.aps.org/prd/pdf/10.1103/PhysRevD.88.104023}{Phys. Rev. D {\bf 88}, 104023} (2013), 	\href{https://arxiv.org/abs/1304.2858}{arXiv:1304.2858 [gr-qc]}

\bibitem{rotating-atom}
It could also be relevant from experimental point of view, along the lines of recent proposal in:
K. Lochan, H. Ulbricht, A. Vinante, S. Goyal,
\href{https://journals.aps.org/prl/abstract/10.1103/PhysRevLett.125.241301}{Phys. Rev. Lett. {\bf 125}, 241301} (2020)
\href{https://arxiv.org/abs/1909.09396}{arXiv:1909.09396}





%
\end{thebibliography}
\end{document}